# Beam Learning - Using Machine Learning for Finding Beam Directions

Saidhiraj Amuru

(Abstract) Beamforming is the key enabler for wireless communications in the mmWave bands. 802.11ad and WiGig are wireless technologies that currently use the 60GHz unlicensed mmWave spectrum via beamforming techniques. It is likely that 5G systems will be considered for 60GHz unlicensed spectrum (apart from other unlicensed bands) deployments and hence must co-exist with 802.11ad and WiGig. 3GPP is taking steps towards achieving the same and the standardization for this is underway. The first step to achieve this co-existence is to find the interference-free directions, in other words identify the directions in which the nodes using these incumbent technologies are communicating and eliminate those directions from further communications. Such a mechanism can help to exploit the spatial holes rather than avoid communications even when only a few spatial directions are used by incumbents. Such a mechanism trivially increases the throughput of the proposed 5G systems.

However, since the incumbent technologies may be unknown to the 5G mmWave nodes and their behavior may also be unknown a priori (for instance, parameters such as duty cycle, power levels, CSMA parameter used by 802.11ad are unknown to the 5G nodes), this spatial direction finding must be performed in a blind manner. In this paper, we use multi-armed bandits-based algorithms, a variant of machine learning algorithms, to blindly detect the beam directions (both along azimuth and elevation i.e., 3D-beamforming) used for communication by the incumbents. This work paves the way for combining the powerfulness of machine learning algorithms into 5G unlicensed mmWave systems. Numerical results show the superior performance of these algorithms over techniques that are commonly employed in such blind settings.

## 1. INTRODUCTION

Large bandwidth available in the millimeter wave (mmWave) spectrum is attractive for making 5G systems a reality [1]. Owing to the channel propagation characteristics in this spectrum, large antenna arrays are typically employed to perform beamforming and transmit data in a directed manner. Note that large antenna arrays are typically employed by both the base station (BS) and the user equipment (UE) to provide directional gains for both transmission and reception operations which thereby help in compensating for the high path loss and high channel absorption seen in these bands [1].

Since traditional beamforming, which is done at the baseband level, requires one RF chain per transmitting antenna (highly complex hardware and power in-efficient), hybrid-beamforming which is a combination of analog and digital beamforming, is being explored to make mmWave communications a reality [2]-[4]. Analog beamforming is usually implemented by using phase shifters and is much simpler than the digital beamforming. In this paper, our focus is on the analog beamforming aspects of the mmWave systems.

Another recent technological advance for enabling mobile traffic offloading is the operation of the cellular technologies in the unlicensed bands. For example, LTE-Unlicensed and licensed-assisted-access (LAA) are the recent technologies studied by various stakeholders, including 3GPP, to analyze the potential benefits of LTE operations in the unlicensed bands [5]. While co-existence studies are still being done across various groups in 3GPP, preliminary results and analysis indeed show that LTE can co-exist well with Wi-Fi in the 5GHz unlicensed band [5].

3GPP and various other entities such as METIS, have started studying the feasibility of 5G deployments and operations in the unlicensed 3.5GHz, 5GHz and 60GHz bands. Considering the successful design phase of LAA in 3GPP, it is envisioned that the 5G unlicensed system design in 3.5GHz and 5GHz bands can leverage off the LTE unlicensed design considered in the 5GHz band. However, the 60GHz unlicensed band is a mmWave band and behaves much differently compared to the sub-6GHz spectrum bands. WiGig and 802.11ad are the existing technologies that operate in this spectrum band which operate in a directional manner based on beamforming (as opposed to the traditional omni-directional based wireless communications used in cellular technologies so far). Considering these facts, new design requirements and considerations should be taken into account for the unlicensed band 5G operations in the 60 GHz spectrum.

The first step for co-existing with the incumbents in the 60GHz spectrum is to identify the interference-free beam directions i.e., identify the spatial directions used by the incumbents for their communication. However, this has to be done in a real-time setting and in a blind manner since the direc-tional information may not be available *a priori* to the 5G mmWave nodes. The simplest technique is to sense in an omni-directional manner and decide whether or not there is any energy in the vicinity. But due to the channel characteristics of the mmWave bands, the coverage of omni-based sensing techniques is severely limited (the coverage of such a transmission-reception scheme is limited). In order to improve this, a sector-based sensing mechanism as in WiGig and 802.11ad may be used [6], for example by employing a specialized hardware namely the electronically steerable parasitic array radiator. However, this technique requires offline non-convex optimizations to be performed before it can be deployed in real-time environments. Hence, it is not truly an online technique. Furthermore, the number of sectors scanned by this technique severely limits the performance as will be shown in this paper. Some enhancements were considered in [7], [8] where a fixed number of beams/ antenna states were used to find the best beam directions. Again, as we will show in this paper, having a fixed number of beams/antenna states *a priori* is sub-optimal in terms of the sensing performance as these fixed beam directions may not be the ones used by the incumbent nodes. Also note that while wider beams fasten the sensing process, narrower beams have a better sensing performance but at increased complexity. However, since the number of beams cannot be decided without *a priori* knowledge about the type of the interferer, the existing techniques fail to work well in practical environments.

In this paper, we develop novel multi-armed bandit (MAB) based algorithms for beam learning i.e., learning the beam directions used by the incumbents. *In contrast to the existing works, see [6]-[9] and references therein, we consider the case of* 3D-*beamforming which is one of the key-enablers for full-dimension MIMO in* 4G *systems and also* 5G *systems* [10]. With the advent of software-defined radios, it is very likely that such machine learning-based algorithms can be deployed in real environments. Furthermore, the proposed algorithms can provide performance bounds as well on the beam learning performance indicating how fast the interference-free beam directions can be identified. Specifically, our algorithms show how to adapt the azimuth and the elevation angles of the array response vectors, both of which are continuous variables. This introduces significant challenges as opposed to the existing works which study a much simpler problem of learning the beam directions using a linear array and by assuming only a fixed number of beam directions i.e. discrete parameter learning. Numerical results show the superior performance of the proposed algorithms in blind settings. Therefore, the main contributions of this paper are

1. the first blind-learning algorithm for learning the beam directions in a 3D-beamforming setting, and
2. theoretical bounds on the performance of the proposed learning algorithm.

At a high-level, in the proposed algorithm, the UE first decides on the beam directions that it must scan to identify potential neighbors i.e., BS-user pairs. Upon decoding the beam directions, it can scan these directions in an intelligent manner, specifically by using the exploration-exploitation dilemma, and then decide the beam directions it can use freely with high guarantee of no interference. The challenges include identifying the beam directions and the intelligent scanning methods. These will be discussed in detail in this paper.

The rest of the paper is organized as follows. We introduce the system model in Section 2. The bandit-based algorithms used for beam learning are introduced in Section 3, where we develop novel beam learning algorithms and present regret[1] bounds for the associated learning performance. Numerical results are presented in Section 4 where we study the behavior of the learning algorithms and compare their performance against commonly used techniques in blind settings and finally conclude the paper in Section 5.

*Notation*: The following notations are used in this paper. $\mathbf{X}$, $\mathbf{x}$ and $x$ indicate a matrix, vector and a scalar respectively. $\mathbf{x}^H$ and $\mathbf{x}^T$ indicate the Hermitian transpose and the transpose operations respectively. $\mathbf{I}$ indicates an identity matrix, $\mathcal{N}(\mathbf{m}, \mathbf{\Sigma})$ indicates a complex Gaussian random vector with mean $\mathbf{m}$ and co-variance matrix $\mathbf{\Sigma}$. $\mathbb{E}[\mathbf{x}]$ is used to indicate the expectation operator, $||\mathbf{x}||$ is the Frobenius norm of $\mathbf{x}$ and finally $|x|$ indicates the absolute value of x .

## 2. SYSTEM MODEL AND PROBLEM FORMULATION

Consider the mmWave system shown in Fig. 1. The system is modeled along the lines of the mmWave system considered in [11]. We consider an 802.11ad or WiGig or any other technology node that exists in the spectrum of interest and has $N_{BS}$ antennas and communicates with its user as shown in Fig. 1 using the red-blue beam pair. For ease of exposition we term these nodes as the base station (BS) and user respectively. Consider a node that intends to operate in the same spectrum and is equipped with $N_{UE}$ antennas with its beams shown in orange color in Fig. 1. For ease of exposition, we term this node as the user equipment (UE). For both UE and BS, we assume that they are equipped with a 2D-planar antenna array (much like most advanced wireless technologies such as massive MIMO, 5G mmWave etc.) with the respective number of antennas i.e., $N_{BS}$ and $N_{UE}$. Under such a system setting, the UE intends to find the directions used by the BS to communicate with its user.

Let $s$ denote the transmitted symbol from BS to its user such that $\mathbb{E}[|s|^2] = P$, where $P$ is the average total transmitted power by BS. Let $\mathbf{f_{RF}}$ indicate the $N_{BS} \times 1$ analog beamforming vector used by the BS towards its user i.e., to form a directional beam towards its user. The samples of the overall transmitted signal is given by

$$\mathbf{x} = \mathbf{f_{RF}} s. \quad (1)$$

$\mathbf{f_{RF}}$ is the analog beamforming matrix which is implemented using phase-shifters. Therefore, the matrix entries are unit-modulus and $\mathbf{f_{RF}}$ is normalized such that $||\mathbf{f_{RF}}||^2 = 1$.

Since we are interested in the UE's ability to find the directions used by the BS and its users, we next focus on the signals received by the UE. A narrow band block-fading chan-nel model is assumed in this paper [11]-[13]. Therefore, the received signal at the UE is given by,

$$\mathbf{r} = \mathbf{H}\mathbf{f_{RF}}s + \mathbf{n}, \qquad (2)$$

where $\mathbf{n} \sim \mathcal{N}(0, \sigma^2 \mathbf{I})$ is the additive white Gaussian noise at the UE, $\mathbf{H}$ is the $N_{UE} \times N_{BS}$ channel between the BS and the UE. Since the UE has a planar antenna array, $\mathbf{r}$ and $\mathbf{n}$ indicate the vectorized versions of the received signal and noise respectively.

---

[1] In the current context, regret is defined as the difference between the cumulative reward of the optimal strategy used by the UE when there is complete knowledge about the beam directions, and the cumulative reward achieved by the proposed learning algorithm.

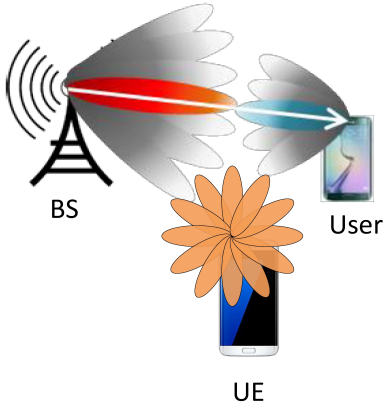

**Fig. 1.** The figure shows that the BS and its user use the red-blue beam pair for their communication. The UE scans the channel using its beams shown in orange color in an attempt to find the directions used by the BS and its users.

In order to identify the BS signals and the directions used, the UE must tune its array response vector to the corresponding directions used by the BS. Let $\mathbf{w}_{UE}$ denote the UE's array response vector. Therefore, the signal used by the UE for processing and identifying the directions is given by,

$$y = \mathbf{w}_{UE}^H \mathbf{H}\mathbf{f_{RF}}s + \mathbf{w}_{UE}^H \mathbf{n}, \qquad (3)$$

where $\mathbf{w}_{UE}$ has similar constraints as $\{\mathbf{f}_i^{RF}\}_{i=1}^{U}$. The focus of this work is on finding the best $\mathbf{w}_{UE}$ in a blind and real-time online setting so that it can find interference-free dimensions for its own usage.

We use the geometric-channel model in [14] with $L$ scatterers. Specifically, the mmWave channel between the BS and UE can be modeled using a ray-based model with $L$ paths as

$$\mathbf{H} = \sqrt{\frac{N_{BS}N_{UE}}{L}} \sum_{l=1}^{L} \alpha_l \mathbf{a}_{UE}(\theta_{UE,l}, \phi_{UE,l}) \mathbf{a}_{BS}^H(\theta_{BS,l}, \phi_{BS,l}), \qquad (4)$$

where $\alpha_l$ is the complex fading coefficient of the $l$th path including the path-loss, with $E(|\alpha_l|^2) = \bar{\alpha}$. The variables $\theta_{UE,l}$ and $\theta_{BS,l}$ indicate the azimuth angles of arrival and departure at UE and BS respectively for the $l$th path. Similarly, $\phi_{UE,l}$ and $\phi_{BS,l}$ indicate the elevation angles of arrival and departure at UE and BS respectively for the $l$th path. Finally, $\mathbf{a}_{BS}(\theta_{BS,l}, \phi_{BS,l})$ and $\mathbf{a}_{UE}(\theta_{UE,l}, \phi_{UE,l})$ are the array response vectors for the planar antennas at the BS and UE respectively. For instance, the BS array response vector [14] is given by (5) shown at the top of the next page. $\mathbf{a}_{UE}$ is formed along the same lines.

In (5), $Y_{BS}$ and $Z_{BS}$ are the number of antenna elements along the horizontal and vertical dimensions of the planar array of BS such that $Y_{BS} * Z_{BS} = N_{BS}$ and $0 \leq m < Y_{BS}$, $0 \leq n < Z_{BS}$, $d$ is the inter-element spacing and $\lambda$ is the wavelength of the spectrum under consideration. Due to the propagation characteristics of the mmWave bands, $L << \min(N_{BS}, N_{UE})$. This is because mmWave channels are by nature low delay spread channels due to high attenuation and narrow beamforming.

Note that $\mathbf{a}_{UE}$ is unknown to the UE as it is operating in a blind setting. Therefore, the objective of this paper is to find $\mathbf{w}_{UE}$ in an efficient manner such that it matches $\mathbf{a}_{UE}$ without any *a priori* knowledge. In this paper, we use the received signal SNR as the metric for finding the directions used by the BS and its user. Note that RSRP is a typical metric used in 3GPP as well for identifying the best beam pairs, beam link failures etc. Hence, using SNR as the metric in this study is well justified.

## 3. BEAM LEARNING

In this section, we propose online learning algorithms to learn the available beam directions for communications from UE perspective. In other words, the proposed algorithms help the UE to avoid the beam directions used by the BS and its users.

### 3.1. Set of actions for the UE

In order to learn the beam directions, the UE must
1. tune its array response vector to a particular direction,
2. receive signals from the BS (if at all transmitted in that direction) and evaluate SNR along this direction and
3. conclude whether or not this direction is used by the BS and its user.

The 2nd step above can be replaced by more elaborate techniques such as decoding the preamble (which is typically known in the case of open access technologies such as 802.11ad and others) and evaluating the associated error rate. However, in such cases the procedure must be completed within the time duration where preambles are transmitted. Due to this restriction, in this paper, we rely on SNR at the UE to take a decision on the beam directions, which is given by

$$\text{SNR} = \frac{P}{\sigma^2}|\mathbf{w_{UE}}^H(\theta_{UE}, \phi_{UE})\mathbf{H}\mathbf{f_{RF}}|^2. \qquad (6)$$

**Remark 1.** The goal of this paper is to show how bandit algorithms can be used in blind settings for identifying the beam directions and thereby minimize the regret which is defined with SNR as the metric. The analysis in this paper can be extended to consider the error rate metric in a straightforward manner by borrowing some analysis from [16].

Irrespective of the metric used for analysis, the UE must learn the array response vector $\mathbf{w}_{UE}$ which is given by (7) shown at the top of the next page. In (7), $Y_{UE}$ and $Z_{UE}$ are the number of antenna elements along the horizontal and vertical dimensions of the planar array of UE such that $Y_{UE} * Z_{UE} = N_{UE}$ and $0 \leq m' < Y_{UE}$, $0 \leq n' < Z_{UE}$, Since $d$ and

$$\mathbf{a}_{BS}(\theta_{BS,l}, \phi_{BS,l}) = \frac{1}{\sqrt{N_{BS}}} \left[ 1, \ldots, \exp\left(j\frac{2\pi}{\lambda} d(m \sin(\theta_{BS,l})\sin(\phi_{BS,l}) + n\cos(\phi_{BS,l}))\right), \ldots, \right.$$
$$\left. \exp\left(j\frac{2\pi}{\lambda} d((Y_{BS}-1)\sin(\theta_{BS,l})\sin(\phi_{BS,l}) + (Z_{BS}-1)\cos(\phi_{BS,l}))\right) \right]^T \quad (5)$$

$$\mathbf{w}_{UE}(\theta_{UE}, \phi_{UE}) = \frac{1}{\sqrt{N_{UE}}} \left[ 1, \ldots, \exp\left(j\frac{2\pi}{\lambda} d(m' \sin(\theta_{UE})\sin(\phi_{UE}) + n'\cos(\phi_{UE}))\right), \ldots, \right.$$
$$\left. \exp\left(j\frac{2\pi}{\lambda} d((Y_{UE}-1)\sin(\theta_{UE})\sin(\phi_{UE}) + (Z_{UE}-1)\cos(\phi_{UE}))\right) \right]^T. \quad (7)$$

$\lambda$ are fixed, the UE has to learn the variables $\theta_{UE}$ and $\phi_{UE}$. Note that, $\theta_{UE} \in [0, 2\pi]$ and $\phi_{UE} \in [-\frac{\pi}{2}, \frac{\pi}{2}]$ and hence they are continuous variables. Note that these values can cover the entire cell i.e., in terms of the angular directions of a typical planar array. The UE learns these parameters by first tuning to a specific direction and then evaluating the SNR along this direction. The complete procedure is detailed next.

Note that the UE must take decisions about the directions in a real-time online setting. Since there is not much *a priori* information with the UE, we formulate this decision making problem as a *learning problem*. Specifically, we model it as a continuous multi-armed bandit problem. Different from previous works on identifying beam directions (both learning-based and otherwise) [6]-[9], the action space here consists of continuum (azimuth and elevation angles of arrival) sets of actions. This is a much more challenging problem than traditional multi-armed bandit-based learning problems which typically assume a fixed set of actions that the learner must choose from. Several well known bandit-based algorithms such as UCB1 (proposed in [15]) have been used in the past for addressing such problems as done in [8]. However, in a continuous action setting, such as the ones studied in this paper, it is not straightforward to just discretize the action space with some finite resolution and then using the existing algorithms such as UCB1. As will be shown in this paper, this results in a sub-optimal performance.

To address this problem, in this paper, we intelligently and iteratively perform discretization of the continuous action space and then learn from this resulting set. We also study the performance of other commonly used blind learning algorithms, namely $\epsilon$-greedy and a multi-armed bandit algorithm (upper confidence bound or UCB) proposed in [15] and compare their performance. In all cases, the feedback for the UE for each action (i.e., the azimuth and elevation angle pair) is the associated SNR measured along the chosen direction. These algorithms will also be compared against commonly used blind algorithms.

### 3.2. MAB formulation

The actions (also called the arms) of the MAB are defined by the pair $\{\theta_{UE}, \phi_{UE}\}$. The strategy set $\mathcal{S}$ that constitutes the azimuth and elevation angles of arrival (that will be used by the UE, which is a subset of $[-\pi/2, \pi/2] \times [0, 2\pi]$ where $\times$ indicates the Cartesian product), is a compact subset of $\mathbb{R}^2$, which is the 2-dimensional real space. For each time $t$, a cost function (feedback metric) $C_t : \mathcal{S} \to \mathbb{R}$ is evaluated. Since we are interested in finding the beam directions used, we define $C_t = \text{SNR}_t$ which is the SNR of the signals received by the UE at time $t$ when using a particular strategy $\mathbf{s}_t \in \mathcal{S}$. This cost function needs to be learned by the UE over time in order to optimize its beam scanning strategy.

Since the action set is a continuum of arms, it is assumed that the arms that are close to each other (in terms of the Euclidean distance), yield similar expected costs. Such assumptions on the cost function will at least help in learning strategies that are close to the optimal strategy (in terms of the achievable cost function) if not the optimal strategy, especially when we consider learning continuous parameters [18]. Formally, the expected or average cost function $\bar{C}(\mathbf{s}) : \mathcal{S} \to \mathbb{R}$ is assumed to be uniformly locally Hölder continuous with constant $L_H \in [0, \infty)$ and exponent $\alpha_H \in (0, 1]$. More specifically, the Hölder condition is given by

$$|\bar{C}(\mathbf{s}) - \bar{C}(\mathbf{s}')| \le L_H ||\mathbf{s} - \mathbf{s}'||^{\alpha_H}, \quad (8)$$

for all $\mathbf{s}, \mathbf{s}' \in \mathcal{S}$ with $||\mathbf{s} - \mathbf{s}'|| \le \delta > 0$ [16]. Here, $||\mathbf{s}||$ denotes the Euclidean norm of the vector $\mathbf{s}$. We assume that the UE knows $L_H$ and $\alpha_H$. The next theorem shows that this similarity assumption holds true when the cost function is SNR.

**Proposition 1.** SNR is uniformly locally Hölder continuous.

*Proof*: The SNR evaluated along a particular direction used by the UE is given by

$$\text{SNR} = \frac{P}{\sigma^2} |\mathbf{w_{UE}}^H(\theta_{UE}, \phi_{UE}) \mathbf{H} \mathbf{f_{RF}}|^2. \quad (9)$$

For two-strategies $\mathbf{w_{UE,1}}$ and $\mathbf{w_{UE,2}}$ used by the UE, we

have

$$|\text{SNR}_1 - \text{SNR}_2| = \\ |\mathbf{w_{UE,1}}^H \mathbf{HH}^H \mathbf{w_{UE,1}} - \mathbf{w_{UE,2}}^H \mathbf{HH}^H \mathbf{w_{UE,2}}| \quad (10)$$

For $\mathbf{w_{UE,1}}$ and $\mathbf{w_{UE,2}}$ such that $\mathbf{s}_1 = \{\theta_{UE,1}, \phi_{UE,1}\}$ and $\mathbf{s}_2 = \{\theta_{UE,2}, \phi_{UE,2}\}$ satisfy $||\mathbf{s}_1 - \mathbf{s}_2|| \leq \delta$, then it is clear that SNR is uniformly locally Hölder continuous for some $L_H$ and $\alpha_H$. This is because for small changes in the values of $\theta_{UE}$ and $\phi_{UE}$ the received SNR does not change much so long as the changes are much smaller than the beam resolution $\frac{2}{\sqrt{N_{UE}}}$. This is typically true for the case of practical beam forming designs which assume a brick-wall type beam forming as within the beam resolution, not much changes in SNR is seen. This was also observed by performing several simulations. Also, note that the exact values of $L_H$ and $\alpha_H$ depend on the antenna configurations. As noted in [19], even if the UE does not exactly know $L_H$ and $\alpha_H$, some bounds on these values will work well.

### 3.3. Proposed Algorithm

In this paper, we propose a novel algorithm for learning the beam directions used by the BS with its users which is termed as the *Beam Learning Bandits* algorithm. The Beam Learning Bandits (BLB) is shown in Alg. 1. At each time $t$, it forms an estimate $\hat{C}_t$ on the cost function $\bar{C}$, which is an average of the costs observed over the first $t-1$ time slots. However, since the actions are continuous, the algorithm discretizes them and then approximately learns the cost function among these discretized versions. For example, $\theta_{UE}$ is discretized as $2\pi\{1/M, 2/M, \ldots, 1\}$ and $\phi_{UE}$ is discretized as $\frac{-\pi}{2} + \pi\{1/M, 2/M, \ldots, 1\}$, where $M$ is the *discretization* parameter. The value of $M$ is derived based on the regret bounds for the BLB algorithm.

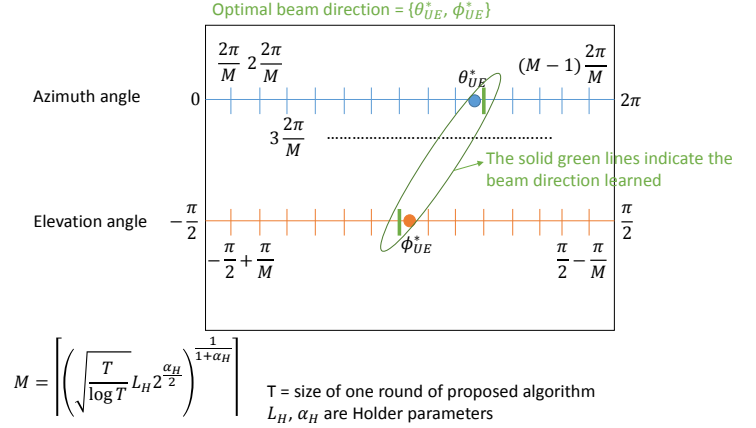

Fig. 2. An illustration of learning in one round of BLB. It is possible that the optimal strategy denoted by $\{\theta_{UE}^*, \phi_{UE}^*\}$ lies out of the set of discretized strategies. In such a case the UE learns the best discretized strategy, but based on the value of the discretization parameter $M$, the loss incurred by using this strategy with respect to the optimal strategy can be bounded using the Hölder continuity condition. The value of the discretization $M$ is shown in the figure and Alg. 1.

---

**Algorithm 1** Beam Learning Bandits (BLB)

T← 1
1: **while** $T \leq n$ **do**
2: $\quad M \leftarrow \lceil (\sqrt{\frac{T}{\log T}} L_H 2^{\alpha_H/2})^{\frac{1}{1+\alpha_H}} \rceil$, $M = 1$ if $T = 1$
3: $\quad$ Initialize UCB1 algorithm [15] with strategy set $2\pi\{1/M, 2/M, \ldots, 1\} \times -\pi/2 + \pi\{1/M, 2/M, \ldots, 1\}$, where $\times$ indicates the Cartesian product.
4: $\quad$ **for** $t = T, T+1, \ldots, \min(2T-1, n)$ **do**
5: $\quad\quad$ Get strategy $\mathbf{s}_t$ from UCB1 [15]
6: $\quad\quad$ Play $\mathbf{s}_t$ and receive the feedback $C_t(\mathbf{s}_t)$ i.e., estimate the SNR at time $t$
7: $\quad\quad$ For each arm in the strategy set, update its index using $C_t(\mathbf{s}_t)$.
8: $\quad$ **end for**
9: $\quad T \leftarrow 2T$
10: **end while**

---

Alg. 1 divides the entire time horizon $n$ into several rounds with different durations. Within every round (the duration $T$ of each round is also adaptive as shown in Alg. 1), it uses a different discretization parameter $M$ to create the discretized joint action set, and learns the best beam direction over this set. The operations of BLB in one such round is shown in Fig. 2. The discretization $M$ increases with the number of rounds as a function of $T$. Its value given in line 2 of Algorithm 1 balances the loss incurred due to exploring actions in the discretized set and the loss incurred due to the sub-optimality resulting from the discretization. In other words, it is chosen such that the regret incurred by BLB is minimized. The various losses incurred and the derivation of the optimal value for $M$ will be explained in detail in Theorem 1. In summary, upon discretization of the continuous arm space, the UE chooses the azimuth and elevation angles for its beam directions by using the UCB1 algorithm shown in Algorithm 2, which is a well known multi-armed bandit algorithm [15]. Therefore, the outer loop of the algorithm adaptively changes the time duration of the inner loop and provides it with the discretization parameter $M$ while the inner loop performs discretization of the arm space and chooses the best arm among these discretized arms by using UCB1. A high-level summary of the proposed algorithm is shown in Fig. 3.

---

**Algorithm 2** Upper confidence bound-based MAB algorithm - UCB1

**Initialization**: Play each arm once
**Loop**:
Scan the beam along $\theta_{UE}, \phi_{UE}$, which maximizes $\hat{C}(\underbrace{\theta_{UE}, \phi_{UE}}_{\mathbf{s}}) + \sqrt{\frac{2\log t}{u_{\mathbf{s}}}}$ where $t$ is the time duration since the start of the algorithm, $u_{\mathbf{s}}$ is the number of times the arm $\mathbf{s} = \{\theta_{UE}, \phi_{UE}\}$ has been played and $\hat{C}(\underbrace{\theta_{UE}, \phi_{UE}}_{\mathbf{s}})$ is the estimated average reward obtained from this arm.

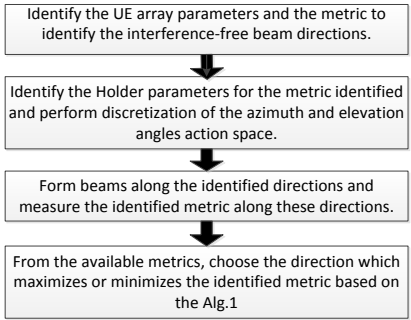

**Fig. 3**. A high-level summary of Alg. 1.

Note that BLB does not need to know the exact time horizon $n$. Time horizon $n$ is only given as an input to BLB to indicate the stopping time. All our results in this paper hold true for any time horizon $n$. This is achieved by increasing the time duration of the inner loop to $2T$ at the end of every round (popularly known as the doubling trick [18]). Another stopping criteria could include comparison of the rewards being obtained over a period of time and UE makes a decision whether the rewards have converged or not. Such a decision can be left to the UE. In this paper, we run the algorithm for a fixed and long time duration assuming that the rewards will converge within this time duration. In Alg. 1, the inner loop can use any of the standard finite-armed MAB algorithms such as UCB1 [15], which is shown in Algorithm 2 for completeness.

### 3.4. Upper bound on the regret

The $n$-step regret $R_n$ is the expected difference in the total cost between the strategies chosen by the proposed algorithm i.e., $\{\mathbf{s}_1, \mathbf{s}_2, \ldots, \mathbf{s}_n\}$ and the best strategy $\mathbf{s}^*$. More specifically, we have $R_n = \mathbf{E}\Big[\sum_{t=1}^{n} C_t(\mathbf{s}^*) - C_t(\mathbf{s}_t)\Big]$, where the expectation is taken over the random feedback signals.

**Theorem 1.** The regret of **Beam Learning Bandits** is $\mathcal{O}(n^{\frac{\alpha_H+2}{2(\alpha_H+1)}} (\log n)^{\frac{\alpha_H}{2(\alpha_H+1)}})$.

*Proof*: The proof of the Theorem is based on the Hölder continuity properties of the cost function established in Proposition 1. The proof is shown in the Appendix.

**Remark 2.** The regret decreases as $\alpha_H$ increases because higher values of $\alpha_H$ indicate that it is easier to separate strategies (i.e., in the current context beam directions) that are close (in Euclidean distance) to each other.

**Corollary 1.** The average cumulative regret of BLB converges to $0$. Its convergence rate is given as $\mathcal{O}(n^{\frac{-\alpha_H}{2(\alpha_H+1)}} (\log n)^{\frac{\alpha_H}{2(\alpha_H+1)}})$.

*Proof*: The proof follows in a straightforward manner from Theorem 1.

See that the average cumulative regret converges to $0$ as $n$ increases. These results establish the learning performance i.e., the rate of learning (how fast the regret converges to $0$) of BLB and indicate the speed at which the beam directions are learned by the UE using Algorithm 1. Since the proposed algorithms and hence their regret bounds are dependent only on $L_H$ and $\alpha_H$, which are in turn a function of the antenna parameters of the UE, BS, etc., the proposed algorithms can be extended to a wide variety of wireless settings by only changing these parameters. The exact values of $L_H$ and $\alpha_H$ need not be known in these cases (because the UE may not have complete knowledge of the wireless channel, BS and its user), the worst case $L_H$ and $\alpha_H$ can be used in the proposed BLB algorithm. This will be shown in the Section 4 where we show the performance of the proposed algorithms.

### 3.5. Discussion

CKL-UCB is a recently proposed continuous armed bandit algorithm [17]. "CKL-UCB explores the apparently suboptimal arms by choosing the least played arms first" [17]. The rationale behind CKL-UCB is that, given a set of suboptimal arms, by exploring them, first eliminate the arms whose expected reward is low (also, these arms do not require many plays to be eliminated). Furthermore, for the case of a single continuous parameter learning, CKL-UCB uses the following discretization $\lceil \left(\sqrt{\frac{T}{\log T}}\right) \rceil$ which does not consider the behavior of the feedback metric (as done in BLB) used for learning. For more details on the implementation of CKL-UCB, please see [17].

Several assumptions are made in the development of CKL-UCB. For instance, CKL-UCB assumes a Lipschitz similarity metric while our model considers and proves the applicability of the Hölder similarity metric condition, which is a more general continuity assumption and can be applicable to many cost functions that may arise in wireless settings. Furthermore, as opposed to BLB, no specific regret bounds are given for CKL-UCB when it is used for learning continuous parameters. In addition, reward distributions belonging to a certain parametrized family of distributions are necessary for CKL-UCB to work. However, no such assumptions are necessary for BLB to work in real-time settings.

The computational complexity of CKL-UCB is significantly more when compared to BLB since the arm to play at a given time instant is chosen by first performing Euclidean distance calculations between every pair of arms, and mutual information calculations for each arm and then by comparing these values across all arms [17]. The goal of the algorithms in [17] was to achieve regret-bound optimality. However, we would like to mention that we do not claim any such optimality for BLB with respect to the regret bounds achieved (although performance wise, it indeed reaches the optimal performance as will be shown in the numerical results). Our goal is only to provide practically feasible algorithms that can be used in real-time wireless settings for beam learning.

## 4. NUMERICAL RESULTS

We assume that the antenna elements are critically spaced $d = \frac{\lambda}{2}$. For such a setting, it is clear that the algorithms proposed in this paper work for any frequency band that needs beamforming for communication. The coefficients $\alpha_l$ for the $L$ paths are generated according to a Rayleigh fading model such that $\bar{\alpha} = 1$. The noise variance $\sigma^2 = 1$ and hence SNR $= \frac{P}{\sigma^2}$. Since mmWave channels are typically sparse, the value of $L$ used in this paper is limited to 5 [11]. The parameters in BLB are set as $L_H = 4$ and $\alpha_H = 1$ (as mentioned earlier, only upper bounds on these values can work well for BLB). We set the azimuth and elevation angles of arrival at the UE for the 1st path are both $\frac{\pi}{3}$. So, the UE must learn these values without *a priori* knowledge. All other angles (departure angles and arrival angles for the various multi path components) are generated randomly.

In the results below, we show the performance of the $\epsilon$-greedy learning algorithm with $\epsilon = 0.9$ and the UCB1 algorithm. These are the most basic learning algorithms used in blind settings [15], even in several wireless communication settings. For $\epsilon$-greedy, we use an exponentially decreasing exploration schedule for this algorithm where the exploration probability decreases as $\epsilon^{\frac{t}{10}}$. For both UCB1 and $\epsilon$-greedy, the actions to try are generated based on discretization of the action space i.e., the azimuth and elevation angles, with a resolution of 5, 10 and 20. In other words, 5, 10 or 20 values that are uniformly spaced between the minimum and maximum values of the elevation and azimuth angles are chosen and tried by the UE for learning the beam directions. Note that the action space for these algorithms is formed from the Cartesian product of the $M$ actions of azimuth angle and $M$ actions of the elevation angle. Using these algorithms independently for each of these action spaces results in a very poor performance and hence is not studied in this paper [23].

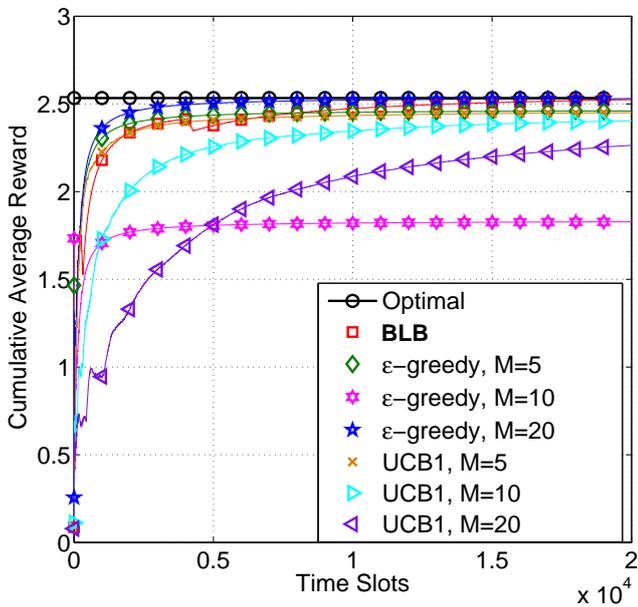

**Fig. 4. Beam learning performance of various algorithms, $N_{BS} = 64, N_{UE} = 16$.**

Fig. 4 shows the performance of the various algorithms when $SNR = -20dB$ and $N_{BS} = 64, N_{UE} = 16$. We plot the cumulative average reward (in terms of the SNR) over time that is seen at the UE. As expected, the $\epsilon$-greedy and UCB1 algorithms with different discretization levels perform differently. Their performance varies based on the action space chosen. Only when the optimal values of actions (azimuth and elevation angles) lies within the discretized space, then their performance will be comparable to the optimal performance. For instance, while the performance of $\epsilon$-greedy with discretization of 20 is close to the optimal performance and also to that of BLB, the performance of the other discretization values is not good. Similar behavior is seen in the case of the UCB1 algorithm, albeit with a different discretization value. This behavior clearly indicates that assuming a specific discretization values does not always guarantee the best performance in a real-time online setting. On the other hand, the proposed BLB algorithm manages to reach the optimal performance by iteratively learning the optimal action choices. This indicates that the regret of the proposed algorithm converges to 0 as shown earlier in Corollary 1.

Fig. 5 shows shows the performance of the various algorithms when $N_{BS} = 64, N_{UE} = 64$. See that the gap between the performance of BLB and other algorithms is now larger as compared to Fig. 4. This is because, as the number of antenna array elements increases, the beam resolution becomes finer and the overlap between various actions decreases (overlap defined in terms of the directions covered by various beam directions scanned using the azimuth and elevation angle actions in the learning algorithms), thereby showing a larger difference in the rewards obtained between different arms. This difference is only expected to increase as the number of antenna elements increases further. Even in Fig. 5 it is seen that the performance of BLB is close to the optimal value and that the performance of $\epsilon$-greedy and UCB1 algorithms is highly dependent on the discretization value chosen.

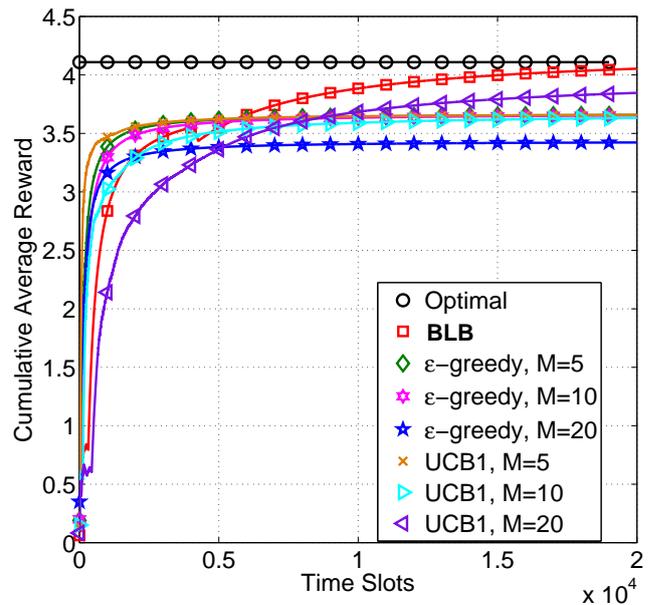

**Fig. 5. Beam learning performance of various algorithms, $N_{BS} = 64, N_{UE} = 64$.**

### 4.1. Learning performance against a BS-User pair with time varying beam directions

Here, we consider the performance of the proposed beam learning algorithm when the BS and its user changes their directions across time. For ease, we assume that only the elevation angle changes over time and the azimuth angle remains fixed. The performance of BLB when the BS-user pair employs such a time varying strategy is shown in Fig. 6. When the beam direction changes rapidly, BLB cannot track the changes perfectly as seen in Fig. 6 because it learns over all the past information, and prior information may not convey knowledge about the current beam directions used by the BS-user pair which can be completely different from the prior beam directions. Specifically, in Fig 6, we show the elevation angles learned by the UE under such a time varying scenario. It can be seen that the UE is not able to learn and adjust to the beam directions used by the BS-user pair. In such cases, it is important to learn only from recent past history, which can be achieved by using BLB on a recent window of past history (for instance, a sliding window-based algorithm to track changes in the environment) [24]. Specifically, we use the concept of drifting [24] to adapt to the victim's strategy.

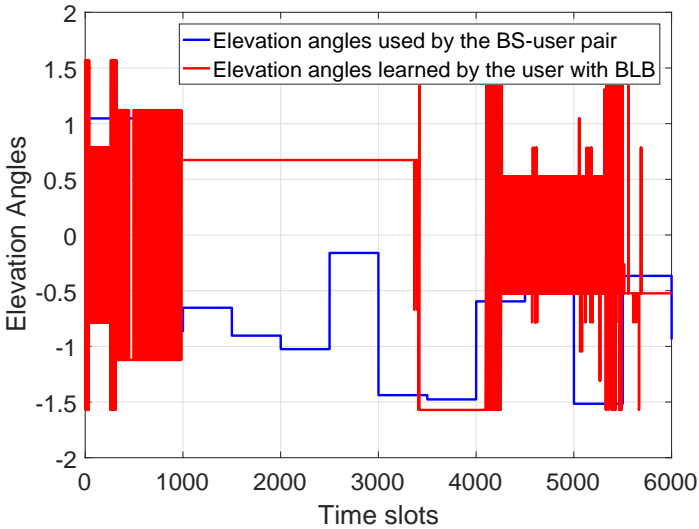

**Fig. 6.** Learning against a BS-user pair with time-varying beam directions. The figure shows the elevation angle (in radians) adaptation by the UE when it uses Alg. 1 $N_{BS} = 64, N_{UE} = 64$.

In this algorithm, each round $i$ (which is of $T$ time steps, where $T = 2^i$) is divided it into several frames each of $W$ time instants. Within each frame, the first $W/2$ time steps, are termed as the passive slot and the second $W/2$ time instants are termed as the active slot. In the first frame, both the slots will be taken to be active slots. Each passive slot overlaps with the active slot of the previous frame. If time $t$ belongs to active slot of frame $w$, then actions are taken as per the UCB1 indices evaluated in this particular frame $w$. However, if it belongs to the passive slot of frame $w$, which is taken to overlap with the active slot of frame $w-1$, then it takes actions as per the UCB1 indices of the frame $w - 1$, but updates the UCB1 indices so that it can be used in frame $w$. Specifically, at the start of every frame $w$, the counters and mean reward estimates are all reset to zero and when actions are taken in the passive slot of frame $w$, these counters and reward estimated are updated so as to be used in the active slot. Thus when the algorithm enters the active slot of frame $w$, it already has some observations using which it can exploit without wasting time in the exploration phase. Such splitting of the time horizon will enable the UE to quickly adapt to the BS-user's varying strategies. Please see [24] for more details on the drifting algorithm. Specifically, we consider the drifting algorithm with a window length $W = 250$.

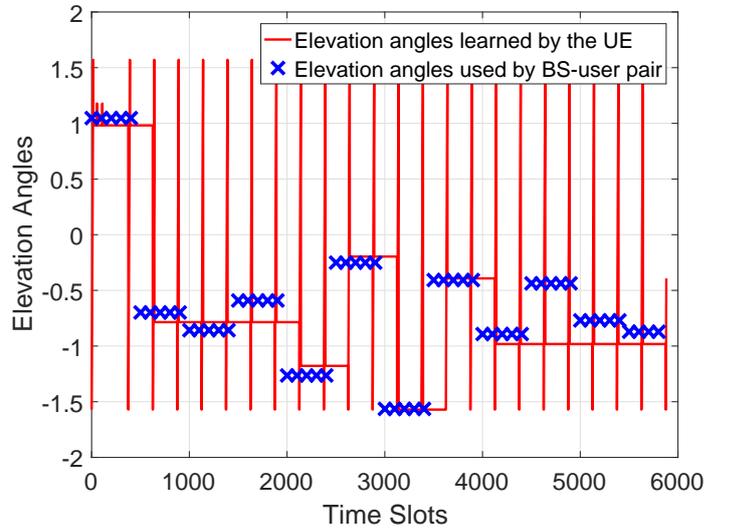

**Fig. 7.** Learning against a BS-user pair with time-varying beam directions. The figure shows the elevation angle (in radians) adaptation by the UE when it uses a drifting algorithm, $N_{BS} = 64, N_{UE} = 64$.

Fig. 7 shows the elevation angles (in radians) learned by the UE when the BS-user pair changes its directions in a time varying manner. Here, we assumed that the elevation angles randomly changes with a uniform distribution in $[0, 2\pi]$ and the azimuth angle stays constant. The dips seen at regular intervals in Fig. 7 are due to the proposed sliding window-based algorithm where the user resets the algorithm at regular intervals to adapt to the changing wireless environment. After these dips, it can be seen that the beam directions learned by the UE is converged to some value (horizontal red lines) which indicates that the UE is indeed capable of tracking the actual beam directions used by the BS-user pair. While Fig. 7 considered the case when the BS-User pair changes its beam directions randomly, the UE can also easily track the beam directions when the BS-user pair employ commonly used adaption strategies such as adapting to interference etc. These results successfully illustrate the adaptive capabilities of the proposed learning algorithms that can overcome the difficulties faced by BLB as shown in Fig. 6.

### 4.2. Impact to 5G Systems

Providing high data rates is one of the KPIs for upcoming 5G systems. In the case of mmWave unlicensed bands, it is therefore crucial to maximize the gains by utilizing any freely available direction for communications. As indicated in the numerical results, the proposed algorithms can successfully identify the freely available directions in a 5G mmWave unlicensed spectrum and thereby allows to improve the spectral efficiency. Further, the proposed algorithms manage to do this in a real-time wireless setting thereby proving their applicability to practical communication settings.

### 4.3. Open Problems

Here, we discuss a few open problems that can be studied for future research.

1. When multiple beam directions are used by the BS to communicate with its users, then the UE must identify each of these directions and avoid them for communication. In the proposed algorithms, the UE chooses the directions based on the best SNR received from the directions it has scanned. The same problem arises wherein multiple different unlicensed technologies are present in the vicinity of the UE and the UE must be capable of identifying all these directions. The received signal at the UE for such a problem can be represented as

$$\mathbf{r} = \sum_{i=1}^{U} \mathbf{H_i} \mathbf{f_{RF,i}} s_i + \mathbf{n}, \quad (11)$$

wherein $U$ is the total number of unknown nodes/technologies existing in the vicinity of the UE, $\mathbf{H_i}$ is the channel from the $i$th transmitter to the UE which uses a pre-coding vector $\mathbf{f_{RF,i}}$ for sending a signal $s_i$ to its corresponding user. When multiple directions needs to be chosen, the UE must rely on combinatorial-multi-armed bandit algorithms such as the ones used in [22], [23]. Here, the UE must choose the multiple directions based on some optimization problem as done in [22] or based on some simple criteria such as maximization of the products of the rewards obtained along multiple directions as done in [23]. One change that can be done for the proposed Alg. 1 is to not just choose one direction at each time step based on the reward obtained, but choose multiple directions (in this case $U$) at each time step based on the knowledge of the number of existing technologies in its vicinity. However, since a UE can only form one beam at a given time instant, some more intelligent decision must be taken to choose these $U$ directions and needs more deeper investigation.

2. In this paper, we focused only on practically feasible algorithms that may be deployed in real-time wireless settings. However, the optimal bandit algorithms with respect to the regret bounds (such as the ones in [20], [21]) may be explored for wireless communications problems.

3. A practical demonstration of the proposed algorithms will be necessary to prove the feasibility of employing machine learning techniques in such realistic scenarios. Note that these algorithms were indeed implemented in real-time in [8], albeit for a different purpose, and give confidence that these bandit algorithms can be used for practical implementations of wireless communications algorithms.

## 5. CONCLUSION

In this paper, we studied blind beam learning algorithms in practical mmWave networks without having any *a priori* information about the wireless environment. Novel multi-armed bandit based algorithms were proposed to learn the beam directions used by the incumbent technology nodes in a blind setting. The proposed algorithms were capable of learning the optimal beam directions and performed better than the commonly used blind learning algorithms such as $\epsilon$-greedy and UCB1. The gains obtained by the proposed algorithm over commonly used blind algorithms becomes larger as the antenna array size increases because of the increase in the array beam resolution. Moreover, the proposed algorithm comes with theoretical guarantees on the learning performance in terms of the regret incurred during the learning process. The proposed algorithm can therefore be used for enabling real-time co-existence of various beamforming-based mmWave systems in the unlicensed spectrum. We also present another version of the algorithm wherein a concept called drifting is borrowed from the existing literature and is applied to time-varying settings wherein the BS-user pair change their communication directions randomly. We showed that even in this setting, the proposed algorithm can closely track the directions used by the BS-user pair. These evidences prove the applicability of the proposed algorithms in a wide variety of realistic wireless settings.

## APPENDIX
## PROOF OF THEOREM 1

Since the time horizon of the inner loop of Algorithm 1 is $T$, we first show that the regret incurred by the inner loop is $\mathcal{O}(\sqrt{M^2 T \log(T)})$. Since the overall time horizon is generally unknown, the algorithm is run for several rounds of time steps on the order of $2^i$ as shown in Algorithm 1 and the regret bounds for the overall algorithm can be achieved by using the doubling trick [25].

The upper bound on the overall regret incurred by Algorithm 1 can be obtained by upper bounding $\sum_{t=1}^{T} \left( \bar{C}(\mathbf{s}^*) - \bar{C}(\mathbf{s}_t) \right)$, where $\bar{C}$ indicates the average cost function and $\mathbf{s}^*$ is the best strategy i.e., optimal direction to be found by the UE which in other words is the direction used by the BS and its user for communicating; and $\mathbf{s}_t$ is the actual strategy chosen at time $t$. We obtain the regret bound in two steps by rewriting it as

$$\sum_{t=1}^{T} \left( \bar{C}(\mathbf{s}^*) - \bar{C}(\mathbf{s}_t) \right) = \sum_{t=1}^{T} \left( \bar{C}(\mathbf{s}^*) - \bar{C}(\mathbf{s}') \right)$$
$$+ \sum_{t=1}^{T} \left( \bar{C}(\mathbf{s}') - \bar{C}(\mathbf{s}_t) \right), \quad (12)$$

where $\mathbf{s}' \in 2\pi\{1/M, 2/M, \ldots, 1\} \times \{-\pi/2 + \pi\{1/M, 2/M, \ldots, 1\}\}$ is the strategy nearest (in terms of the Euclidean distance) to $\mathbf{s}^*$. Then we have $||\mathbf{s}' - \mathbf{s}^*|| \leq \sqrt{\frac{2}{M^2}}$ based on the discretization of the continuous arms set in Algorithm 1. For the first term in the above equation, by using the Hölder continuity properties of the average cost function $\bar{C}$, we have

$$\mathbf{E}\Big( \sum_{t=1}^{T} C_t(\mathbf{s}^*) - C_t(\mathbf{s}') \Big) = \sum_{t=1}^{T} \left( \bar{C}(\mathbf{s}^*) - \bar{C}(\mathbf{s}') \right)$$
$$\leq TL\Big(\frac{2}{M^2}\Big)^{\alpha/2}. \quad (13)$$

We now bound the second term $\mathbf{E}\Big( \sum_{t=1}^{T} C_t(\mathbf{s}') - C_t(\mathbf{s}_t) \Big) = \sum_{t=1}^{T} \left( \bar{C}(\mathbf{s}') - \bar{C}(\mathbf{s}_t) \right)$. Due to the discretization technique used in Algorithm 1, this problem is equivalent to a standard MAB problem with $M^2$ arms [15]. In order to bound (13), we define two sets of arms: near-optimal arms and sub-optimal arms. We set $\Delta = \sqrt{M^2 \log(T)/T}$ and say that an arm is sub-optimal in this case, if its regret incurred is greater than $\Delta$ and near-optimal when its regret is less than $\Delta$. Thus, for a near-optimal arm, even when that arm is selected at all time steps, the contribution to regret will be at most $T\Delta$. In contrast for a sub-optimal arm, the contribution to the regret when it is selected can be large. Since we use the UCB1 algorithm, it can be shown that the sub-optimal arms will be chosen only $\mathcal{O}(\log(T)/\Delta(\mathbf{s})^2)$ times ($\Delta(\mathbf{s})$ is the regret of the strategy $\mathbf{s}$) [15], before they are identified as sub-optimal. Thus the regret for these sub-optimal arms is on the order of $\mathcal{O}(\log(T)/\Delta)$ since $\Delta(\mathbf{s}) > \Delta$. From these arguments the second term in (12) can be upper bounded as

$$\mathbf{E}\Big( \sum_{t=1}^{T} C_t(\mathbf{s}_t) - C_t(\mathbf{s}') \Big) \leq \mathcal{O}(\sqrt{M^2 T \log(T)}). \quad (14)$$

Using (13) and (14), and setting $M = \lceil (\sqrt{\frac{T}{\log(T)}} L 2^{\alpha/2})^{\frac{1}{1+\alpha}} \rceil$ (this is obtained by matching the regret bounds shown in (13) and (14)), the regret for any given signaling scheme is given by $\mathcal{O}(\sqrt{M^2 T \log(T)})$. By using the value of $M$, the doubling trick, and summing the regret over all inner loop iterations of Algorithm 1, the regret over the entire time horizon $n$ can be expressed as $\mathcal{O}(n^{\frac{\alpha_H+2}{2(\alpha_H+1)}} (\log n)^{\frac{\alpha_H}{2(\alpha_H+1)}})$.